\documentclass[%
 reprint,
 superscriptaddress,
 nofootinbib,
 amsmath,
 amssymb,
 aps,
 floatfix,
]{revtex4-1}

\usepackage{color}
\usepackage[T1]{fontenc}
\usepackage{graphicx}
\usepackage{dcolumn}
\usepackage{bm}
\usepackage{hyperref}

\newtheorem{thm}{Theorem}

\begin{document}


\title{Information estimation using nonparametric copulas}

\author{Houman Safaai} \email{houman\_safaai@hms.harvard.edu.} 
 \affiliation{Department of Neurobiology, Harvard Medical School, Boston, MA}
 \affiliation{Istituto Italiano di Tecnologia, Rovereto, Italy}
\author{Arno Onken}
\affiliation{School of Informatics, University of Edinburgh, Edinburgh, UK}
\author{Christopher D. Harvey}
 \affiliation{Department of Neurobiology, Harvard Medical School, Boston, MA}
\author{Stefano Panzeri}
\affiliation{Istituto Italiano di Tecnologia, Rovereto, Italy}

\date{\today}

\begin{abstract}
Estimation of mutual information between random variables has become crucial in a range of fields, from physics to neuroscience to finance. Estimating information accurately over a wide range of conditions relies on the development of flexible methods to describe statistical dependencies among variables, without imposing potentially invalid assumptions on the data. Such methods are needed in cases that lack prior knowledge of their statistical properties and that have limited sample numbers. Here we propose a powerful and generally applicable information estimator based on non-parametric copulas. This estimator, called the non-parametric copula-based estimator (NPC), is tailored to take into account detailed stochastic relationships in the data independently of the data's marginal distributions. The NPC estimator can be used both for continuous and discrete numerical variables and thus provides a single framework for the mutual information estimation of both continuous and discrete data. By extensive validation on artificial samples drawn from various statistical distributions, we found that the NPC estimator compares well against commonly used alternatives. Unlike methods not based on copulas, it allows an estimation of information that is robust to changes of the details of the marginal distributions. Unlike parametric copula methods, it remains accurate regardless of the precise form of the interactions between the variables. In addition, the NPC estimator had accurate information estimates even at low sample numbers, in comparison to alternative estimators. The NPC estimator therefore provides a good balance between general applicability to arbitrarily shaped statistical dependencies in the data and shows accurate and robust performance when working with small sample sizes. We anticipate that the non-parametric copula information estimator will be a powerful tool in estimating mutual information in a broad range of data.

\end{abstract}

\maketitle

\section{\label{sec:intro}Introduction}

Mutual information, the fundamental mathematical quantity of information theory, provides a universal way to quantify dependencies, transmission rates, and representations of data~\cite{Shannon1948}. It has become an indispensable tool in many domains such as signal processing, data compression, finance, dynamical systems, and neuroscience~\cite{Mackay2003, Cover2006, goh2018inference, SPIKES, quiroga2009, Cellucci2005}. Mutual information quantifies the information that one random variable carries about another by measuring the reduction in uncertainty about a given variable from knowing another variable. Uncertainty in turn is quantified by means of entropy. Shannon's entropy therefore is at the core of virtually all applications of information theory.

Quantifying entropy and information of a random variable poses a difficult problem because it requires knowledge about its probability distribution. In most practical applications, the exact shape of the distribution of a random variable is unknown and thus needs to be estimated from data. This requires either strong parametric assumptions, such as assuming for instance that data follow a normal distribution, or large amounts of data to estimate the distribution directly from the samples.

In an ideal case, information estimators would estimate variable distributions directly from the data and would not require parametric assumptions that could impose invalid structures on the data. In addition, ideal estimators would be accurate also in situations with limited sample numbers. Furthermore, given that mutual information quantifies only the dependencies between the variables \cite{Jenison2004,calsaverini2,zeng2011,InceGC}, ideal estimators should be sensitive only to the dependencies between the random variables of interest, which fully define mutual information, and should be insensitive to other aspects of the data, such as the marginal distributions of the individual random variables. To date, it has been challenging to develop information estimators that have all these properties. It is clear that developing such estimators would greatly increase the range of applicability and the accuracy of information measures over a wide range of important empirical problems.

For continuous random variables, powerful estimators have been developed that estimate mutual information directly from the samples in a non-parametric way. One popular class is based on the k-nearest neighbor (kNN) estimators \cite{Kraskov2004,Pal2010,Victor2002}, which in their original form assume local uniformity in the vicinity of each point. For accurate information estimation with these approaches, the required number of samples scales exponentially with the value of mutual information \cite{Gao2015}. This has limited the effectiveness of these estimators in cases with strong dependencies and thus high mutual information, or in situations with smaller numbers of samples. The performance of these estimators, especially for strong dependency cases, has been improved through the introduction of a correction term for local non-uniformity (LNC)\cite{Gao2015}. The LNC method assumes a particular non-uniformity structure of the distribution in the kNN ball or max-norm rectangle. These assumptions, however, can produce inaccuracies in information estimates for data with marginal distributions with long tails, such as the gamma distribution, or distributions with sharp boundaries. Thus, assumptions about local non-uniformity could lead to different estimates of mutual information for two sets of variables that have similar dependency structures, and hence similar mutual information values, but different marginal distributions. These methods therefore encounter a significant trade-off between assumptions imposed on the distribution of the data and the number of samples required for accurate information estimation.

For discrete variables, estimation methods have been proposed based on either subtracting out an analytical approximation to the limited sampling bias~\cite{Paninski2003,Panzeri1996}, or in using a Bayesian approach. In the latter, instead of estimating the probability mass function, a prior, in the form of Dirichlet distributions, is placed over the space of discrete probability distributions. The entropy then is estimated using the inferred posterior distribution over entropy values~\cite{Nemenman2002,Nemenman2004}. A more complete set of priors has been recently proposed in~\cite{Archer2014}, using a mixture of Pitman-Yor priors (PYM), which is a two-parameter generalization of the Dirichlet process and parameterized to be flat over entropy values. It has been shown that such a flat prior provides better estimates of entropy and mutual information with low sample numbers compared to analytical bias subtraction methods~\cite{Nemenman2002,Archer2014}. However, like the LNC estimator, the PYM estimator is sensitive to the form of the marginal distributions. In particular, Gerlach et al.~\cite{Gerlach2016} confirmed that the PYM estimator reduces the estimation bias but that the bias scales in the same way with the number of samples as for other type of estimators. Moreover, the PYM estimator performs worse on heavy-tailed distributions~\cite{Gerlach2016,Wollstadt2017}.

The previously proposed estimators considered above have in common that in one way or another they make use of the full joint distribution of the random variables of interest, which includes contributions from both the marginal distributions and the dependencies between the variables of interest. However, because mutual information is determined only by the dependencies between variables~\cite{Jenison2004,calsaverini2,zeng2011}, information estimators only need to focus on correctly capturing the dependency structure. Such dependency structures are best isolated  using the mathematical construct known as the copula. Formally, any joint distribution can be decomposed into its marginal distributions and a copula. The latter quantifies the dependency structure irrespective of the marginals, and the negative of the copula entropy exactly equals the mutual information that one random variable carries about the other~\cite{Sklar1959, Jenison2004}. Copula-based methods are therefore well suited for isolating the dependencies and are insensitive to the form of marginal distributions. Previous copula based information estimators have been proposed both in the continuous domain \cite{calsaverini2,zeng2011,InceGC} and mixtures between discrete and continuous domains \cite{Onken2016}.  All such copula based information estimators have made use of copulas selected from parametric families. The parametric copula estimators have the advantage of simplicity, but the disadvantage that they make systematic assumptions on the dependency structure of the data~\cite{Jaworski2013}. These assumptions might differ greatly from the real data structure, leading to large estimation errors when used on datasets with complex and non-linear dependency structures that are difficult to fit with simple parametric copula families. However, recently some nonparametric copula estimation methods have been proposed in \cite{Geenens2014,Geenens2017,Nagler2017}, and their properties in density estimation have been studied. Yet, a systematic study of the application of such methods in mutual information estimation is lacking. 

Here, we propose information estimators based on nonparametric copulas (NPCs). These NPC estimators first identify the copula that characterizes the relationship between the random variables of interest and then calculate the entropy of the copula to obtain an estimate of the mutual information. Contrary to parametric copula families, non-parametric copulas do not impose strong assumptions on the shape of the stochastic relationship between the variables of interest and thereby avoid systematic biases in the information estimates. We present methods to identify the copula nonparametrically, both for continuous and discrete data. We show that, compared to previously reported information estimators (in particular the LNC and PYM estimators), NPC mutual information estimators are robust to the parameters of the marginal distribution and perform well in cases with low sample numbers. NPC-based estimators are therefore some of the first information estimators that simultaneously do not impose strong parametric assumptions, can work with relatively small sample sizes, and isolate the dependencies in the data that matter for mutual information both in the continuous, discrete or mixed domains.
\section{\label{sec:cop}Theory and Methodology}
We estimate information by means of copulas and their entropy. Copulas mathematically formalize the concept of statistical dependencies: a given copula quantifies a particular relationship between a set of random variables. Here we give a brief summary of the basics of the copula and its relation to mutual information. We then continue by presenting the non-parametric copula and how it can be computed empirically from given data.

\subsection{\label{sec:formal}Formal copula definition}
A $d$-dimensional copula is the cumulative distribution function $C(u_1,\dots,u_d):\left[0,1\right]^d\rightarrow\left[0,1\right]$ of a random vector defined on the unit hypercube $\left[0,1\right]^d$ with uniform marginals $\mathcal U_{\left[0,1\right]}$ over $\left[0,1\right]$. 
\begin{equation}\label{cop1}
C(u_1,\dots,u_d)=P(U_1\leq u_1,\dots,U_d\leq u_d),
  *\end{equation}
where $U_i\sim\mathcal{U}_{\left[0,1\right]}$.

The great strength of copulas is their utility for representing the statistical relationship between multiple random variables. Copulas can be used to couple arbitrary marginal cumulative distribution functions (CDFs) to form a joint CDF. Sklar's theorem~\cite{Sklar1959, Nelsen2006} lays out the theoretical foundations for this construction:

\begin{thm}{Sklar's theorem$:$}
	For a d-dimensional random vector $\boldsymbol{X}=(X_1,\dots,X_d)$, let $F_{\boldsymbol{X}}$ be its CDF with marginals $F_1$, $\dots$, $F_d$. Then there exists a copula $C$ such that $\forall \boldsymbol{x} \in \mathbb R^d:$
	\begin{equation}
		F_{\boldsymbol{X}}(x_1,\dots,x_d) = C(F_1(x_1),\dots,F_d(x_d)), \,\,\,\,x_i\in\mathbb{R},
	\end{equation}
	$C$ is unique, if the marginals $F_i$ are continuous.
	Conversely, if $C$ is a copula and $F_1$, $\dots$, $F_d$ are CDFs, then the function $F_{\boldsymbol{X}}$ defined by $F_{\boldsymbol{X}}(x_1, \dots, x_d) = C(F_1(x_1), \dots, F_d(x_d))$ is a $d$-dimensional CDF with marginals $F_1$, $\dots$, $F_d$.
\label{thm:sklar}
\end{thm}
Sklar's theorem relates the copula $C$ of Eq.(\ref{cop1}) to the joint distribution function of the variables $U_i=F_i(X_i)$,
\begin{equation}
C(u_1,\dots,u_d)=F_{\boldsymbol{X}}(F_1^{-1}(u_1),\dots,F_d^{-1}(u_d)), \,\,\,\,u_i\in \left[0,1\right]
\end{equation}
where $F_i^{-1}$ are the inverse cumulative distribution functions. For a differentiable copula $C$, we can define the copula probability density function (PDF) $c(u_1,\dots,u_d)=\frac{\partial^d}{\partial u_1\dots\partial u_d}C(u_1,\dots,u_d)$. For $u_i:=F_i(x_i)$ and $f_i(\cdot)$ as PDFs corresponding to the CDFs $F_i(\cdot)$, we can write the copula density as
\begin{eqnarray}\label{cop11}
c(u_1,\dots,u_d)=\frac{f_{\boldsymbol{X}}\left(F_1^{-1}(u_1),\dots,F_d^{-1}(u_d)\right)}{\prod_{i=1}^d f_i(F_i^{-1}(u_i))}.
\end{eqnarray}

This means that the multivariate PDF can be decomposed into the copula density and the product of the marginal densities. The copula can be interpreted as the part of the density function that is independent from the single variable marginals and rather captures the dependencies between the variables. This decomposition is useful to estimate the joint density function and also to estimate the likelihood which is needed in statistical inference, but here in this work we only focus on the copula density as a tool to compute entropy and mutual information.

\begin{figure}[htp!]
  \includegraphics{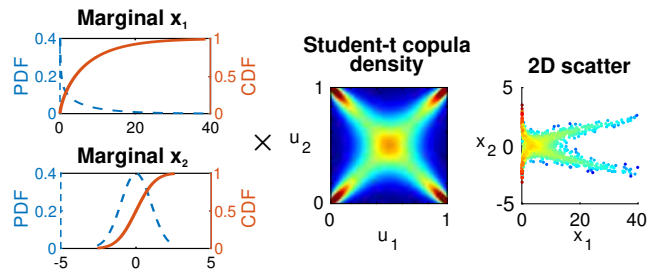}
  \caption{A bivariate dataset (right panel) is generated from mixing two marginal distributions (left panels, gamma top and Gaussian bottom). The PDF (dashed line) and CDF (solid line) for the two marginals are shown. They merge with the copula density (middle panel) to generate the joint density function.}
  \label{schem1}
\end{figure}

An example bivariate density function is shown in Fig.\ref{schem1} which consists of a gamma marginal distribution ($x_1$), a Gaussian marginal distribution ($x_2$), and a particular parametric copula density (student-t copula) as its dependency structure. The decomposition of the full density function into the dependency structure (copula) and the marginal distributions makes it possible to study any measure which is independent from the marginal distributions by considering only the copula structure. Here the gamma marginal distribution has a sharp boundary at $x_1=0$ which makes it difficult for conventional density estimation methods to compute the full bivariate density function. The copula, on the other hand, can easily cope with the density behavior at $x_1=0$.

\subsection{Entropy and mutual information}
\label{sec:information}

Entropy quantifies the uncertainty associated with a given random variable and lays the foundation for mutual information. For a continuous multivariate distribution, the differential entropy $h(\boldsymbol{X})$ is defined as
\begin{equation}
h(\boldsymbol{X}) = -\int f_{\boldsymbol{X}}(\boldsymbol{x})\log_2 f_{\boldsymbol{X}}(\boldsymbol{x})d\boldsymbol{x},
\end{equation}
where $f_{\boldsymbol{X}}$ denotes the multivariate probability density function~\cite{Cover2006, Shannon1948}.
With this, the mutual information $I(\boldsymbol{X};\boldsymbol{Y})$ between two continuous multivariate random variables $\boldsymbol{X}$ and $\boldsymbol{Y}$ is given by
\begin{equation}\label{int1}
I(\boldsymbol{X};\boldsymbol{Y}) = h(\boldsymbol{X}) + h(\boldsymbol{Y}) - h(\boldsymbol{X},\boldsymbol{Y}),
\end{equation}
where $h(\boldsymbol{X},\boldsymbol{Y})$ is the joint differential entropy of the joint distribution $(\boldsymbol{X},\boldsymbol{Y})$ with joint PDF $f_{\boldsymbol{X},\boldsymbol{Y}}$~\cite{Cover2006, Shannon1948}.

Using Eq.(\ref{cop11}), one can show that the mutual information equals the negative of the entropy of the copula density between $\boldsymbol{X}$ and $\boldsymbol{Y}$~\cite{Jenison2004,calsaverini2,zeng2011,InceGC}:
\begin{eqnarray}\label{int2}
I(\boldsymbol{X};\boldsymbol{Y})=-h(c)=\int_{[0,1]^d} c(\boldsymbol{u}) \log_2 c(\boldsymbol{u})\, d\boldsymbol{u},
\end{eqnarray}
where $\boldsymbol{u}=(u_1,\cdots,u_d)$.
This makes the computation of mutual information independent from the marginal distributions and reduces the computational error in estimating the mutual information (MI) for two reasons. First, the irrelevance of the marginals removes the need for faithfully capturing their properties in the information estimation procedure. Thus, copula-based estimators separate the relevant entropy from the irrelevant entropies and thereby effectively reduce the number of implicit quantities contributing to the final mutual information estimate, thereby reducing the estimation error.
Second, the independence of copula from the marginals makes copula based methods robust to any irregularity which might exist in the marginals. This is in contrast to density dependent methods, such as kNN-based estimators \cite{Victor2002,Kraskov2004,Pal2010,Gao2015} which might struggle with marginal irregularities. 

We can estimate the integral Eq.(\ref{int2}) using classical Monte Carlo (MC) sampling~\cite{Robert2004, Onken2016}. The entropy can be expressed as an expectation over the copula density $c$
\begin{equation}
h(c) = - \mathbb{E}_{c}\left[\log_2 c(\boldsymbol{U})\right],
\end{equation}
where $\boldsymbol{U}=(U_1,\dots,U_d)$ denotes a random vector from the copula space. This expectation can then be approximated by the empirical average over a large number of $d$-dimensional samples $\boldsymbol{u}_j=((\boldsymbol{u}_j)_1,\cdots,(\boldsymbol{u}_j)_d)$ from the random vector $\boldsymbol{U}$:
\begin{equation}\label{MC}
-\mathbb{E}_{c}\left[\log_2 c(\boldsymbol{U})\right]  \approx \widehat{h_k} := -\frac{1}{k}\sum_{j=1}^k\log_2(c(\boldsymbol{u}_j))
\end{equation}

By the strong law of large numbers, $\widehat{h_k}$ converges almost surely to $h(c)$. Moreover, we can assess the convergence of $\widehat{h_k}$ by estimating the sample variance of $\widehat{h_k}$: 
\begin{equation}
\text{Var}\left[\widehat{h_k}\right] \approx \frac{1}{k+1}\sum_{j=1}^k\left(\log_2(c(\boldsymbol{u}_j))-\widehat{h_k}\right)^2,
\end{equation}
With this estimate, the term $\frac{\widehat{h_k}-h(c)}{\sqrt{\text{Var}\left[\widehat{h_k}\right]}}$ is approximately standard normal distributed, allowing us to obtain confidence intervals for our differential entropy estimates~\cite{Robert2004}.

\subsubsection{Sampling from the copula\label{sampling}}

To sample from a $d$-dimensional copula, we use the Rosenblatt transform~\cite{Rosenblatt1952, Devroye1986}. This approach applies a sequence of conditional distributions and makes use of the fact that the marginal distributions of a copula are always uniform. First, we draw independent uniform samples $v_1, \dots, v_d$ from $[0,1]$. Then, we sequentially transform these samples by means of the inverse conditional CDFs of the copula:
\begin{eqnarray}\nonumber
u_1 &=& v_1\\\nonumber
u_2 &=& C_{2|1}^{-1}(v_2|u_1)\\\nonumber
u_3 &=& C_{3|1,2}^{-1}(v_3|u_1,u_2)\\\nonumber
&\vdots&\nonumber\\
u_d &=& C_{d|1,\dots,d-1}^{-1}(v_d|u_1,\dots,u_{d-1})
\end{eqnarray}
where $C_{i|1,\dots,i-1}^{-1}$ denotes the inverse of the copula CDF of element $i$ conditioned on the elements $1,\dots,i-1$. The resulting vector $(u_1,\dots,u_d)$ is a sample from the copula.

The conditional CDFs can be obtained from the copula CDF by calculating \cite{Nelsen2006,Joe2015}
\begin{multline}
C_{i|1,\dots,i-1}(u_i|u_1,\dots,u_{i-1}) \\= \frac{\partial^{i-1}C_{1,\dots,i}(u_1,\dots,u_i) / \prod_{k=1}^{i-1}\partial u_k}{c_{1,\dots,i-1}(u_1,\dots,u_{i-1})},
\end{multline}
where $C_{1,\dots,i}$ denotes the copula CDF with the elements $i+1,\dots,d$ marginalized out and $c_{1,\dots,i}$ denotes its PDF.

For the special case $d=2$, computation of the conditional CDF reduces to a partial derivative of the original copula CDF with respect to one variable.

\subsection{Parametric copulas}
\label{sec:parametric}
Many parametric families of copulas have been proposed, representing various relationship shapes with different tail dependencies and  symmetries~\cite{Joe2015,Nelsen2006,Joe1996}. These families are appropriate for fitting data with corresponding features. However, such parametric families make strong assumptions about the shapes of the relationships. This may in turn introduce considerable biases in information estimates when the shape of the dependencies in the real data does not match those that can be described by the copula family.

In this work we will use the parametric copulas for two different purposes. The first is to test the performance of information estimation methods based on parametric copulas. The second is to use particular parametric families to generate data with a known ground-truth information value in order to test the accuracy of our non-parametric copula-based information estimators. For this purpose, the most convenient parametric families are those for which we can analytically calculate mutual information. Two particular parametric families with known closed-form solutions for calculating mutual information are given by the Gaussian and student-t copula families. We describe their properties in this section. For our simulations, we consider only bivariate copulas. However, these copulas can be readily extended to large-dimensional copulas by means of pair-copula-constructions~\cite{Aas2009}, as follows.

\begin{itemize}
\item
\textbf{Gaussian copula family}
One of the most commonly applied parametric copula is the Gaussian copula with CDF defined as $C_{G}(u,v)=\Phi_{\bm{\Sigma}}(\Phi^{-1}(u),\Phi^{-1}(v))$ where $u,v\in\left[0,1\right]$ and $\Phi$ and $\Phi_{\bm{\Sigma}}$ are the univariate standard normal CDF and multivariate normal CDF with zero mean and correlation matrix $\bm{\Sigma} = \begin{pmatrix}1 & r\\ r & 1\end{pmatrix}$, respectively. The copula PDF can be written as
\begin{equation}
c_{G}(u,v)=\frac{1}{\sqrt{|\bm{\Sigma}|}}
\exp\left(-\frac{1}{2}X^\top(\bm{\Sigma}^{-1}-\mathbf{I}_{2})X\right),
\end{equation}
where $\boldsymbol{X}=(x,y)$, $(x,y)=(\Phi^{-1}(u),\Phi^{-1}(v))$ and $\mathbf{I}_2$ denotes the $2\times 2$ identity matrix.

The Gaussian copula entropy has the following analytical form:
\begin{eqnarray}
h(c_G)=-\frac{1}{2}\log_2\left(1-r^2\right),
\end{eqnarray}

\item
\textbf{Student-t copula family}
The student-t copula is another well established parametric copula family which can be used to model elliptical dependency structures. Contrary to Gaussian copulas, copulas from the student-t family have tail dependency and hence can be used to generate datasets with heavy tails. The bivariate student-t copula is defined by means of the standardized bivariate student-t CDF $t_{\bm{\Sigma},\nu}$ as $C_t(u,v)=t_{\bm{\Sigma},\nu}(t_\nu^{-1}(u),t_\nu^{-1}(v))$, where $\bm{\Sigma}$ is the correlation matrix and $\nu$ is the degrees of freedom. The PDF of the bivariate student-t copula is
\begin{equation}
c_t(u,v)=\frac{\Gamma(\frac{\nu+2}{2})\Gamma(\frac{\nu}{2})}{\sqrt{|\bm{\Sigma}|}\Gamma(\frac{\nu+1}{2})}\frac{(1+\boldsymbol{X}^T \bm{\Sigma}^{-1} \boldsymbol{X} / \nu )^{-(\nu+2)/2}}{\left((1+x^2/\nu)(1+y^2/\nu)\right)^{-(\nu+1)/2}},
\end{equation}
where $\boldsymbol{X}=(x,y)$, $(x,y)=(t_\nu^{-1}(u),t_\nu^{-1}(v))$ and $\Gamma(\cdot)$ denotes the gamma function.

The student-t copula has the following analytical entropy \cite{calsaverini2}:
\begin{eqnarray}
h(c_t)=\frac{\Omega}{\ln(2)}-\frac{1}{2}\log_2\left(1-r^2\right),
\end{eqnarray}
where
\begin{eqnarray}
\Omega&=&2\ln\left(\sqrt{\frac{\nu}{2\pi}}\beta\left(\frac{\nu}{2},\frac{1}{2}\right)\right)\\\nonumber
&-&\frac{2+\nu}{\nu}+(1+\nu)\left[\psi\left(\frac{\nu+1}{2}\right)-\psi\left(\frac{\nu}{2}\right)\right]
\end{eqnarray}
is a constant and $\beta(\cdot)$ and $\psi(\cdot)$ are the beta and digamma function, respectively.
\end{itemize}

\subsection{Nonparametric copulas}
Our information estimator is based on a recently developed non-parametric version of the copula, which can be used to model any general dependency structure and does not involve making assumptions on the structure of data \cite{Cellucci2005,Onken2016,InceGC}. One challenge in using non-parametric copula estimators is to deal with the close support of the copula: the support of a bivariate copula is restricted to the unit square $[0,1]^2$. Most kernel estimators, for instance, have problems with such bounded support because for points close to the boundaries, they typically place some positive mass outside of the support. To address this problem, we apply a transformation such that the support of the density in the transformed space is unbounded~\cite{Loader1996, Geenens2014, Geenens2017}.
	
\begin{centering}
\begin{figure}[htp!]
\includegraphics{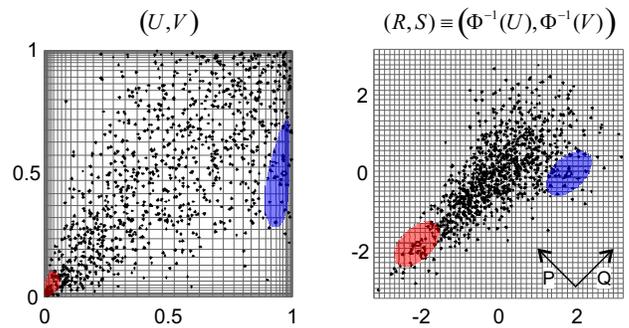}
\caption{$(u_i,v_i)$ samples and their probit transformations $(r_i,s_i)$ are shown. Grids are equally spaced in $(R,S)$ space. The direction of the rotated $(P,Q)$ are shown as insets. The red and blue areas are two example kernel functions.}
\label{bandwfig}
\end{figure}
\end{centering}

Let us assume that we want to estimate a copula density $c$ given $n$ bivariate random samples $(u^i,v^i)$, $i=1\ldots n$ from the random vector $(U,V)$. Let $\Phi$ be the standard normal CDF and $\phi$ its density. Then the random vectors $(R,S)=(\Phi^{-1}(U),\Phi^{-1}(V))$ have normal distributed marginals with support on the full $\mathbb{R}^2$ (Fig.\ref{bandwfig}). In this domain, kernel density estimators work well and have less asymptotic and boundary problems since the density slowly converges to zero on the edges. This transformation is known as the probit transformation.

By sklar's theorem for densities, Eq.(\ref{thm:sklar}), the density $f$ of $(r,s)$ will be decomposed into

\begin{equation} 
f(r,s)=c(u,v)\phi(r)\phi(s).
\end{equation}
After change of variables, we get the copula density
\begin{eqnarray}\label{coptrans}
c(u,v)=\frac{f(\Phi^{-1}(u),\Phi^{-1}(v))}{\phi(\Phi^{-1}(u))\phi(\Phi^{-1}(v))}.
\end{eqnarray}

The non parametric copula can be estimated in several ways, described in what follows.

\subsubsection{Naive kernel estimation}

The naive kernel estimate of the density function can be written as
\begin{equation}
c_\text{naive}(u,v)=\frac{\frac{1}{n}\sum_{i=1}^n K_{\vec b_n}(r-r_i,s-s_i)}{\phi(\Phi^{-1}(u))\phi(\Phi^{-1}(v))},
\end{equation}
where the sum is over the $n$ samples $(r_i,s_i)\equiv (u_i,v_i)$ and $(r,s)$ is related to $(u,v)$ through Eq.(\ref{pcaa}). For the density kernel $K(\cdot)$ we consider a symmetric bounded probability density function with bandwidth vector $\vec b$.
Furthermore, we can make another transformation to the principal component coordinates,
\begin{eqnarray}\label{pcaa}
\binom{p}{q} \equiv \mathbf{W}\binom{r}{s}=\mathbf{W}\binom{\Phi^{-1}(u)}{\Phi^{-1}(v)},
\end{eqnarray}
where the matrix $\mathbf W$ is the rotation matrix to the principal component coordinates. In this coordinate space, since the covariance matrix is diagonal, we can approximate the kernel function as the product of the two kernels for each of the coordinates $K_{\vec b}(p-p_i,q-q_i)\approx K_{b_P}(p-p_i)K_{b_Q}(q-q_i)$ where $b_Q$ and $b_P$ are the corresponding bandwidths of each coordinate. An example of bivariate data is shown both in the $(p,q)$ and $(u,v)$ spaces in Fig.\ref{bandwfig}. 
\subsubsection{Local-likelihood density estimation}
When used for non-parametric copula estimation, the naive kernel estimator has asymptotic problems at the edges of the distribution support. In particular, it might find false peaks and troughs when there is an asymmetry in the tails of the distribution. This happens because small fluctuations in unbalanced tails are greatly magnified when transformed back to the copula space~\cite{Geenens2014}. To remedy this problem, we can make use of a similar approach as in \cite{jones1996}, where it was shown that the the local likelihood density estimation gives a much better behavior on the boundaries~\cite{Geenens2014}. We adapted this approach by assuming that the density function can be written locally for any point $(p',q')$ around each point $(p,q)$ as a continuous function $f(p',q')=\psi_{\bm{\theta}(p,q)}(p-p',q-q')$ for some parameters $\bm{\theta}(p,q)$ and a continuous parametric function $\psi_{\bm{\theta}(p,q)}(p-p',q-q')$. 

The log-likelihood of such an estimate can be written as follows \cite{Loader1996,Geenens2014}
\begin{widetext}
\begin{equation}\label{lik}
\mathcal{L}(p,q)=\frac{1}{n}\sum_{i=1}^n K_{b_p}(p_i-p)K_{b_q}(q_i-q) \log
\psi_{\bm{\theta}(p,q)}(p-p_i,q-q_i)
-\iint_{\mathbb{R}^2} K_{b_p}(p-\tilde p)K_{b_q}(q-\tilde q) \psi_{\bm{\theta}(p,q)}(p-\tilde p,q-\tilde q) d\tilde p\,d\tilde q.
\end{equation}
\end{widetext}
After fixing the functional form for $\psi$, the parameters $\bm{\theta}$ can be obtained by maximizing the log-likelihood
\begin{eqnarray}\label{theta}
\bm{\theta}(p,q)&=&\text{arg}\max\limits_{a_1,\ldots,a_F}   \mathcal{L}(p,q), 
\end{eqnarray}
where we considered $F$ degrees of freedom for $\bm{\theta}$.
A possible choice for the functional form of the $\psi$ studied in \cite{Loader1996,jones1996,Geenens2014,Nagler2016} is to assume that its logarithm is a polynomial. For a polynomial of order 2, the $\psi_{\bm{\theta}}$ around each point $(p,q)$ can be written as
\begin{equation}\label{ppsi}
\psi_{\bm{\theta}}(p-p',q-q')=a_1 e^{a_2 (p-p')+a_3 (q-q')+a_4 (p-p')^2+a_5 (q-q')^2},
\end{equation}
where $\bm{\theta}=(a _1,a_2,a_3,a_4,a_5)$ are the parameters to be defined at each point $(p,q)$. Note that the local likelihood density function is equal to $f_{\text{LL}}(p,q)=a_1(p,q)$. This particular functional form simply means that locally and not globally, around each point $(p,q)$, the log-likelihood function has a Gaussian form. The choice of the kernel functions $K(\cdot)$ are of lower importance, since they will be weighted with the local function $\psi$. Given that the data in the probit coordinates is normal, and has diagonal covariance matrix, the Gaussian kernel function seems to be a natural choice,
\begin{eqnarray}\label{kerng}
K_b(u)=\frac{1}{\sqrt{2\pi}b} e^{-\frac{u^2}{2b^2}},
\end{eqnarray}
 We can now solve Eq.(\ref{theta}) by imposing $\frac{\delta \mathcal{L}(p,q)}{\delta \bm{\theta}}=0$ and solving the following set of equations which we get after using Eqs.(\ref{theta}) and (\ref{lik}) at each point $(p,q)$:
\begin{eqnarray}\label{loklikeq}
\nonumber
 \begin{pmatrix} f_{\text{naive}}\\ f_1 \\ f_2 \\ f_3 \\ f_4  \end{pmatrix} && := \frac{1}{n}\sum_{i=1}^n \begin{pmatrix} 1\\(p_i-p)\\(q_i-q)\\(p_i-p)^2\\(q_i-q)^2\end{pmatrix} K_{b_p}(p_i-p)K_{b_q}(q_i-q)  
 \\\nonumber
 && =   \iint_{\mathbb{R}^2} \begin{pmatrix} 1\\\tilde p \\ \tilde q\\\tilde p^2\\\tilde q^2 \end{pmatrix} K_{b_p}(\tilde p)K_{b_q}(\tilde q)  \psi_{\bm{\theta}(p,q)}(\tilde p,\tilde q) d\tilde pd\tilde q
\\
\end{eqnarray}
The set of equations Eqs.(\ref{loklikeq}) can be solved analytically for the Gaussian kernel as follows:
\begin{eqnarray}\label{fseq}\nonumber
\begin{pmatrix}
f_{\text{naive}}\\f_1\\f_2\\f_3\\f_4
\end{pmatrix}   
=
\frac{a_1}{e_p e_q} 
\begin{pmatrix}
1 \\ \frac{a_2 b_p^2}{e_p^2} \\ \frac{a_3 b_q^2}{e_q^2} \\ \frac{b_p^4 a_2^2+b_p^2 e_p^2}{e_p^4} \\ \frac{b_q^4 a_3^2+b_q^2 e_q^2}{e_q^4}
\end{pmatrix}\exp\left(\frac{a_2^2 b_p^2}{2 e_p^2}+\frac{a_3^2 b_q^2}{2 e_q^2}\right)
\\
\end{eqnarray}
where $e_p$ and $e_q$ are defined as $e_p:=\sqrt{1-2b_p^2 a_4}$ and $e_q:=\sqrt{1-2b_q^2 a_5}$.

The functions $f_{\text{naive}}, f_1, f_2, f_3, f_4$ can be computed empirically for given bandwidths $b_p$ and $b_q$ and from the summation over the data points $(p_i,q_i)$, with $i=1,\cdots,n$, in Eq.(\ref{loklikeq}). We can then solve for the likelihood-estimated copula density $f_{\text{LL}}(p,q)$ as $f_{\text{LL}}(p,q)=a_1(p,q)$ using the following identities which can be extracted from Eqs.(\ref{fseq}):
\begin{eqnarray}
\begin{aligned}[l]
&a_2=\frac{e_p^2}{b_p^2}\frac{ f_1(p,q)}{ f_{\text{naive}}(p,q)}, a_3=\frac{e_q^2}{b_q^2}\frac{ f_2(p,q)}{ f_{\text{naive}}(p,q)}
\\
&e_p =b_p \left( \frac{ f_3(p,q)}{ f_{\text{naive}}(p,q)} - \left(\frac{ f_1(p,q)}{ f_{\text{naive}}(p,q)}\right)^2 \right)^\frac{1}{2}
\\
&e_q =b_q \left( \frac{ f_4(p,q)}{ f_{\text{naive}}(p,q)} - \left(\frac{ f_2(p,q)}{ f_{\text{naive}}(p,q)}\right)^2 \right)^\frac{1}{2},
\end{aligned}
\end{eqnarray}
which can be used to compute the local-likelihood copula density at each point $(p,q)$ as
\begin{eqnarray}\label{copulAn}
f_{\text{LL}}&&(p,q) =
\\\nonumber
&&f_{\text{naive}} e_p e_q \exp\left(-\frac {e_p^2 }{2 b_p^2}\left(\frac{ f_1}{ f_{\text{naive}}}\right)^2-\frac{e_q^2 }{2  b_q^2}\left(\frac{ f_2}{ f_{\text{naive}}}\right)^2\right).
\end{eqnarray}
The copula density function Eq.(\ref{copulAn}) can be computed at any point using Eqs.(\ref{fseq}). This equation gives an analytic correction to the naive density estimate $f_{\text{naive}}$ for the local-likelihood density $f_{\text{LL}}$. The only unknowns at this point are the kernel bandwidths $b_p$ and $b_q$ which will be discussed in the next section.

After computing the density in the $(p,q)$ space, we can transform back to the probit dimensions $(r,s)$ and then to the original $(u,v)$ in the CDF domain $u,v \in [0,1]$
\begin{eqnarray}
\binom{r}{s} = \mathbf{W}^{-1}\binom{p}{q}, \quad  \binom{u}{v} =  \binom{\Phi(r)}{\Phi(s)},
\end{eqnarray}
where $\mathbf W$ is the transformation matrix to the PCA coordinates. The transformation from $(p,q)$ to $(r,s)$ is an isometry, hence $f_{\text{LL}}(r(p,q),s(p,q))=f_{\text{LL}}(p,q)$. The copula density is then computed using Eq.(\ref{coptrans}).

Selecting proper bandwidths is crucial to get well behaved and precise kernel density estimates specially on the borders. This is a sensitive issue which can drastically affect the local and asymptotic properties of the density estimation. The transformation of the data to probit coordinates and then to the principal components makes it natural to consider a diagonal bandwidth matrix as we did in the previous section with two diagonal components $b_r$ and $b_s$ as the only remaining parameters which should be estimated in Eq.(\ref{copulAn}).

There are two main approaches for estimating the bandwidths. In the first one, we consider a constant bandwidth for all the points on the plane while in the second one, we define the bandwidth according to the local distribution of the data, for example to be proportional to the distance of each point to its $k^\text{th}$-nearest neighbor point. Since we want to take advantage of the analytical solution for the local-likelihood copula, we here use a fixed bandwidth.

As discussed in \cite{Geenens2017,Geenens2014}, a good choice of the bandwidth should balance the integrated asymptotic squared bias and the variance of the considered estimator. We do this by minimizing the mean integrated squared error (MISE). However, popular data-driven selection strategies are based on cross-validation. The most popular instances are least-squares cross-validation~\cite{Rudemo1982} and biased cross-validation~\cite{Scott1987}. Here, the MISE takes the form
\begin{eqnarray}\label{MISE}
\text{MISE}[f_{\text{LL}}]&=& \iint_{\mathbb{R}^2} \mathbb{E} \left[ \left( f_{\text{LL}}(p,q)-f_T(p,q)\right)^2 \right] dp dq
\nonumber \\
&\propto&\iint_{\mathbb{R}^2} f_{\text{LL}}^2(p,q) dp dq-\frac{2}{n}\sum_{i=1}^n f_{\text{LL}}^{\{CV\}}(p_i,q_i),
\nonumber\\
\end{eqnarray}
where $f_T$ is the true density of the data points $(p_i,q_i)$ and $f_{\text{LL}}$ is the local likelihood approximate of the density. The $\sum_{i=1}^nf_{\text{LL}}^{\{CV\}}$ term is the cross-validated sum of the copula density over the data points. For instance, a leave-one-out or k-fold procedure can be used to split the data into training and test subsets. The density at each test set can be estimated using the density function which is estimated using only the corresponding training set. By having a cross-validated copula density for each point, we can estimate the sum in the second term. The integral part is computed numerically using the equally spaced binning of the $(p,q)$ space as it is shown in Fig.\ref{schem1}. The possible effect of the number of bins on the mutual information estimation will be shown in the next sections.

The bandwidth parameters $\bm{b}=(b_p,b_q)$ can then be estimated numerically by minimizing the $\text{MISE}[f_{\text{LL}}]$
\begin{equation}\label{bandwid}
\bm{b}=\text{arg}\min\limits_{b_p,b_q} \left\{  \iint_{\mathbb{R}^2} f_{\text{LL}}^2(p,q) dp dq-\frac{2}{n}\sum_{i=1}^n f_{\text{LL}}^{\{CV\}}(p_i,q_i) \right\}.
\end{equation}
In the results presented in this paper we used 5-fold cross-validation to estimate MISE. We did not observe any significant difference in results when using values of $k$ as large as $20$, for a dataset with $n=1000$ samples.

\subsubsection{Bandwidth selection}\label{sec:bandwidth}
One possible simplification for the bandwidth selection is used in \cite{Nagler2016,Geenens2017} where it was shown that a \textit{rule of thumb} way of defining the bandwidth can perform well. In this rule, the bandwidth will be proportional to the square root of the covariance matrix $\bm{b}\propto \Sigma^{1/2}$ where $\Sigma$ is the empirical covariance matrix in the $(p,q)$ coordinates (so it is diagonal). We can then use this approach and instead of optimizing Eq.(\ref{bandwid}) for two free parameters, we can solve the problem for one parameter $\alpha$ after defining the bandwidth as $\bm{b}=\alpha \Sigma^{1/2}$. We will refer to the density function obtained from the simplified one-parameter bandwidth as LL1 and the density function obtained from two-parameter bandwidth as LL2.
\subsubsection{Copula density normalization}
One important property of the copula density is that it has uniform marginals. It is important to ensure that the estimated empirical copula density satisfies this property as well. This means that we should have
\begin{eqnarray}
\int_0^1 c(u,x)dx=\int_0^1 c(x,v)dx = 1, \quad u,v \in [0,1].
\end{eqnarray}
Because of numerical imperfections and approximations of the kernel estimation, these constraints might be violated. In order to impose these constraints, we follow the iterative normalization suggested in Nagler et al.~\cite{Nagler2016} by repeatedly dividing the copula density by its marginals
\begin{eqnarray}\label{copnorm}
c(u,v) \rightarrow \frac{c(u,v)}{\int_0^1 c(u,x)dx \int_0^1c(x,v)dx}.
\end{eqnarray}
A relatively small number of iterations ($\sim$1000) is sufficient to get copula densities with almost uniform marginals. Finally, in order to be sure that we get a proper density function, we normalize the resulting copula density with its integral over the two-dimensional domain $(u,v)\in [0,1]^2$. This numerical normalization assures that the resulting density satisfies the properties of a copula density.

Also note that numerical computation of the integrals required for the estimation of bandwidths in Eq.(\ref{bandwid}) as well as for the estimation of the density in Eq.(\ref{copulAn}) and in the normalization procedure of Eq.(\ref{copnorm}), we use a grid of the $(p,q)$ domain as shown in Fig.\ref{bandwfig}. In principle, it would be possible to use different grid sizes for bandwidth optimization and for density estimation. For example, it may be useful to use a coarser grid to estimate the bandwidth (since it will be more efficient in terms of computational time) and a finer grid size to estimate the density (to have a higher resolution density estimation) or in the sampling procedure. For the simulations presented in this paper, we used equal grids for bandwidth optimization and density estimation both to get a lower bound of the information estimation error and to simplify the procedure.

\section{Results}
Our approach is to compare the NPC-based estimator with current information estimators using numerical simulations in both continuous and discrete domains. We focus on the performance of the NPC-based estimator in terms of optimizing its parameter selection, and evaluating its accuracy, sensitivity to sample size, and robustness to the form of the marginal distributions. We also focus on comparing the properties of NPC-based estimator to those of the best performing estimators among those currently available. 

\subsection{Continuous variables}
\begin{figure}[htp!]
  \includegraphics{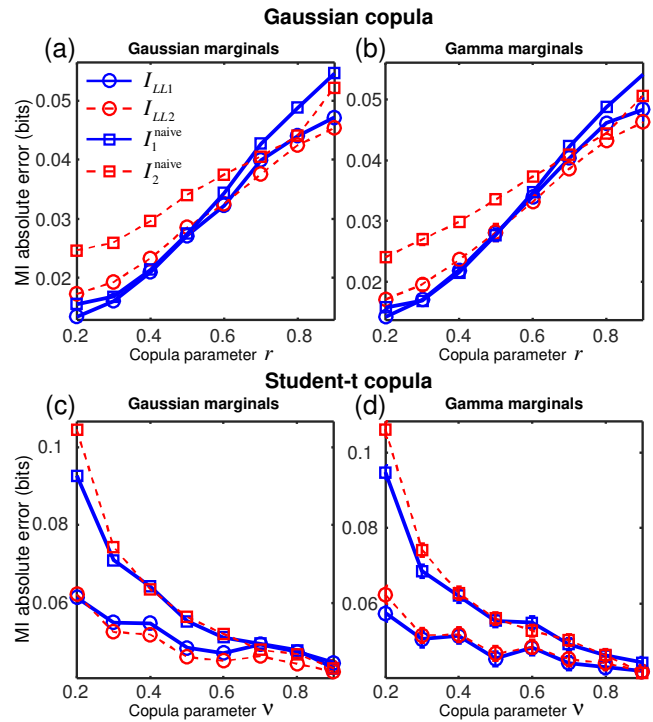}
  \caption{Mutual information absolute error is computed for four different datasets using one-parameter bandwidth ($I_{LL1}$ and $I_1^{\text{naive}}$) and two-parameter bandwidth ($I_{LL2}$ and $I_2^{\text{naive}}$) non-parametric copula using both the naive and local likelihood estimators. Data are generated by means of the copula method with $k=100$ grid size and $N=1024$ sample size. The data in each panel are combinations of two similar marginal distributions and a parametric copula. a) Gaussian copula with parameter $r$, Gaussian marginals. b) Same Gaussian copula, gamma marginals. c) Student-t copula with parameters $r=0$ and $\nu$, Gaussian marginals. d) Same Student-t copula, gamma marginals.}
  \label{naiveI}
\end{figure}
We first consider the case of estimating information between continuous valued variables. These cases are relevant for many important applications, ranging from analysis of gene networks~\cite{margolin2006aracne} to the analysis of neuroimaging data such as electro- and magneto-encephalograms~\cite{InceGC} and to the analysis of continuous valued dynamical systems~\cite{paluvs2003direction,Cellucci2005}.

In the continuous domain, we tested the NPC-based information estimators in four different simulated conditions. We generated the datasets so that we had the ground-truth theoretical values of the mutual information for those probability distributions. We quantified estimation accuracy by computing the mutual information absolute error $E\left[|I_{\text{estimate}}-I_{\text{theory}}|\right]$, the normalized bias $E\left[I_{\text{estimate}}-I_{\text{theory}}\right]/I_{\text{theory}}$ and the normalized variance of the estimator $E\left[(I_{\text{estimate}}-E\left[I_{\text{estimate}}\right])^2\right]/I_{\text{theory}}$ over a number of simulations (1000 simulations for each condition). For each condition, we generated simulated data using a known parametric copula dependency structure and known marginal structures.
For the dependency structure between variables, we considered two families of parametric copulas: the Gaussian copula family and the student-t copula family, each of which has closed-form solutions for calculating the associated entropies (see Section~\ref{sec:parametric}). For the Gaussian copula, $r$ was varied from 0.2 to 0.9. For the Student-t copula, $r$ was set to 0 and $\nu$ was varied between 0.2 and 0.9, forming entirely nonlinear dependencies and zero linear correlation (see the copula in Fig.\ref{schem1}). 

Note that the mutual information is positively correlated with $r$ and negatively correlated with $\nu$. We combined copulas from each of these families with marginal distributions that were either Gaussian $\mathcal{N}(\mu=0,\sigma=1)$ or gamma-exponential $\Gamma(\alpha=0.1,\beta=10)$. The selected parameters for the gamma-exponential marginal distribution formed a sharp boundary peak at zero (similar to the gamma distribution shown in Fig.\ref{schem1}), which is difficult to capture with methods that operate on the properties of the density function. We therefore generated bivariate distributions with selected marginals and a relationship structure specified by the selected copula (see Sklar's theorem~\ref{thm:sklar}). In each case, we simulated the data with the sampling approach explained in Sec.\ref{sampling}. 
\subsubsection{Optimization of the non-parametric copula}
Given that the use of non-parametric copulas has been introduced only recently \cite{chen2007nonparametric,Geenens2014,Racine2015} and that they have not been used for information estimation before, we first investigated how to optimize the performance of various possible implementations of the non-parametric copula (Fig.\ref{naiveI}). We considered versions with a two-parameter bandwidth and a simpler version with a one-parameter bandwidth local-likelihood method (see Sec.\ref{sec:bandwidth}). For all the simulated dataset conditions, we did not find a significant difference between the two- and one-parameter bandwidth versions of the NPC estimator in terms of information estimate accuracy (Fig.\ref{naiveI}). This result suggests that, in the $(p,q)$ space, the covariance of the distribution was enough to capture the local variations in the density and hence the optimal shape of the kernel function. We also compared the absolute mutual information error obtained with the local-likelihood copula with that obtained with the naive copula. It has been already shown that the local-likelihood density copula describes better data with sharp variations, edges or other types of local nonuniformity~\cite{Geenens2014}. 
\begin{figure}[htp!]
  \includegraphics{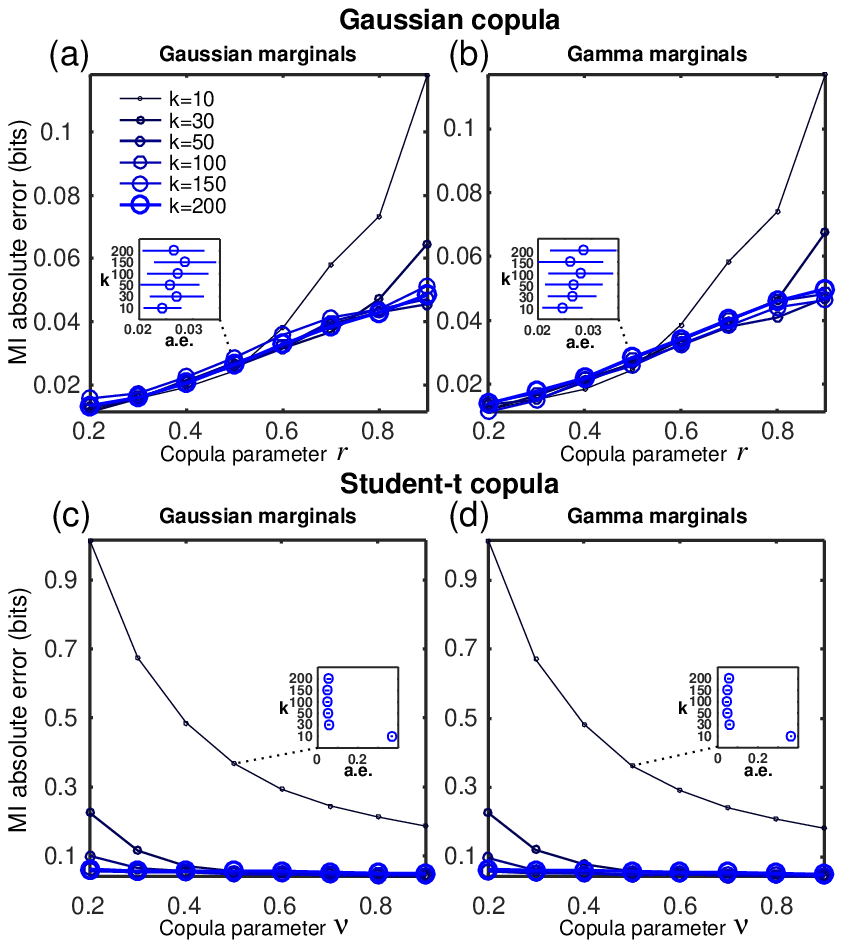}
  \caption{Mutual information absolute error for a range of grid sizes is shown for the same data as of Fig.\ref{naiveI} using the $LL1$ local-likelihood non-parametric copula method. Insets show the $95\%$ confidence interval of the mean of MI absolute error for each $k$ value after correction for multiple comparison for the $r,\nu=0.5$ cases.}.
  \label{contKall}
\end{figure}
In Fig.\ref{naiveI} we tested whether these properties lead to a more accurate mutual information estimation. As expected, the naive estimation of the copula density was accurate in simple situations, such as the Gaussian copula. However, for the case of high non-linear correlations in a student-t copula (smaller $\nu$ values), the naive method failed to capture the sharp corners of the copula and had double the estimation error of the local-likelihood methods. 
We therefore chose to use the $LL1$ as the copula estimator for the comparison with other methods. 

\begin{figure}[hbp!]
  \includegraphics{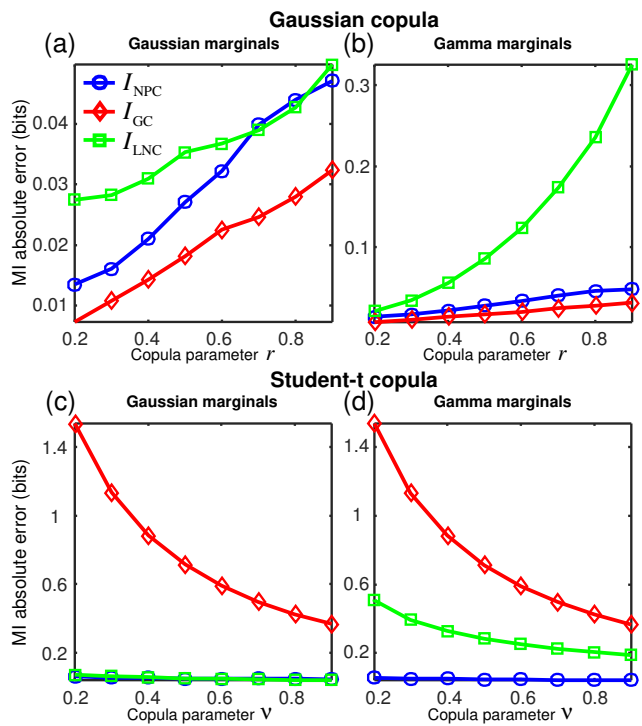}
  \caption{Mutual information absolute error is shown for data similar to Fig.\ref{naiveI} using different information estimation methods.}
  \label{LNCcont}
\end{figure}
\begin{figure}[htp!]
  \includegraphics{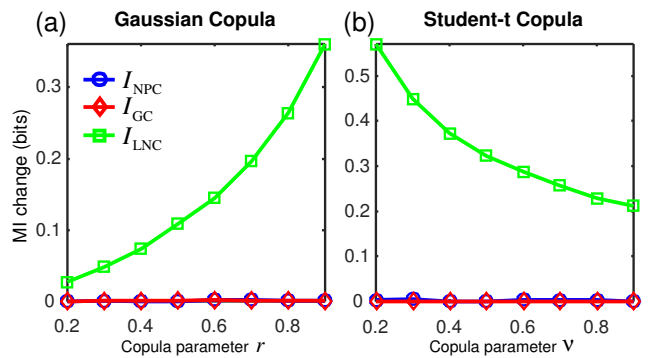}
  \caption{The difference of mutual information estimated with normal and gamma marginal distributions is shown for LNC and NPC methods for (a) the Gaussian copula and (b) the student-t copula, with similar parameters as in Fig.\ref{naiveI}. In all the simulations we used $k=100$ and $N=1024$ samples.}
  \label{LNCdiff}
\end{figure}
This version had the advantage of having fewer parameters for the bandwidth parameterization, which made the optimization of Eq.(\ref{bandwid}) faster and easier to converge, without much cost to the accuracy of the density function and the mutual information estimations.
The only free parameter of the non-parametric copula that needs to be selected a priori is the number of grids $k$ that are used to quantize the $(p,q)$ space for estimation of the bandwidths in Eq.(\ref{bandwid}) and normalization of the copula density in Eq.(\ref{copnorm}). To test how this parameter affected the estimated mutual information, in Fig.\ref{contKall} we tested the NPC estimator, on the same simulated data used in the previous figure, varying $k$ from 10 to 200 (in the previous figure a value of $k= 100$ was used). For $k\ge 50$, there was little improvement in the information error with increasing values of $k$, both in strongly correlated and less correlated copulas. We thus selected $k=100$ for the remaining analysis. For smaller $k$'s, e.g. $k=10$, the resolution of the binning of the copula space was not sufficient to capture the sharp corners of the student-t copula, even though it performed well for the Gaussian copula (Fig.\ref{contKall}). In the practical implementation of the above procedures, we found that, for strongly correlated data (e.g. large $r$ values of the Gaussian copula or small $\nu$ values of the student-t copula), the MI absolute error decreased monotonically with $k$ until reaching a constant value at larger $k$, and a small number of iterations was enough to optimize bandwidth. For weak correlation cases, we still observed a decrease of the estimation error when increasing $k$, although in such cases the copula bandwidths were usually larger and so the bandwidth optimization needed more iterations for large $k$ values. In the results presented in this paper, we used a bounded optimization function since the size of the bandwidth is bounded by the extension of the data in the $(p,q)$ domain.\footnote{The MISE is a convex function which can be optimized easily and reliably. We used the \textsc{Matlab} function \texttt{fminbnd} with maximum 500 number of iterations for the optimization. Using smaller number of iterations as low as 100 will have minimal effect on the results specially in the more correlated cases. The bandwidths are bounded to zero from below and to the rule-of-thumb bandwidth value used in \cite{Nagler2016} from the above.}

\subsubsection{Comparing the NPC estimator in the continuous domain with existing established estimators}
We next compared the NPC method with two other alternative established methods. First, we tested our non-parametric copula estimator against a parametric copula-based estimator based on the Gaussian copula (GC) whose parameters were estimated by maximum likelihood \cite{InceGC}. This estimator was selected for comparison because it is a popular method for estimation of information in continuous brain signals~\cite{InceGC}. Second, we also compared our NPC estimator against the mutual information estimates obtained with the LNC method \cite{Gao2015}. This comparison was chosen because, as we also confirmed in our experience on our simulated data, the LNC method is considered to be the best performing among those not based on copulas such as those based on nearest neighbors ~\cite{Kraskov2004,Victor2002,Gao2015}. The results for all four simulation conditions and for a range of copula parameters are shown in Fig.\ref{LNCcont}. 

The GC gave the most accurate results in the case of data generated using a Gaussian copula, Fig.\ref{LNCcont}, as expected because in this case the parametric copula used for generating the data matched the one used for estimating information, but it gave the largest error in estimating the mutual information on data generated by the student-t copula, which lacked linear correlations in the data. 
\begin{figure}[b!]
  \includegraphics{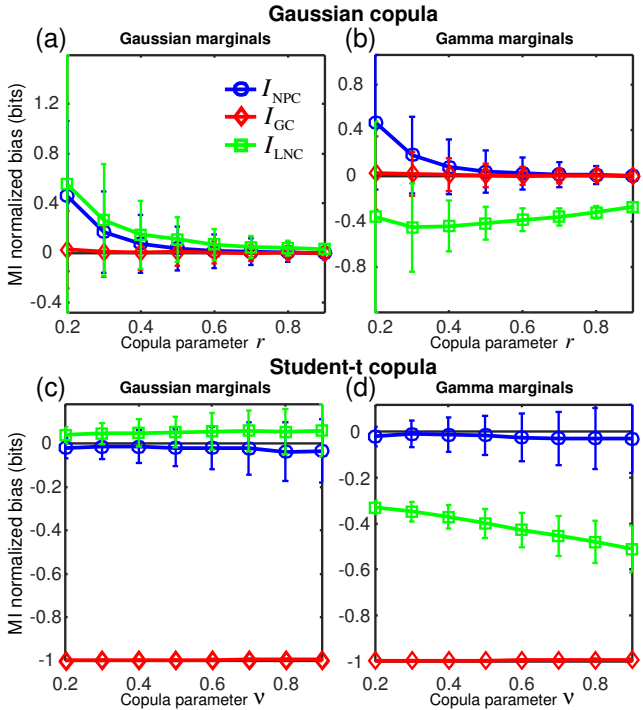}
  \caption{Mutual information bias ratio is shown for similar data as in Fig.\ref{naiveI} using different information estimation methods. The errorbars represent the standard deviation over the 1000 simulations of the data.}
  \label{LNCcont_bias}
\end{figure}

The LNC method worked well for both copula families when we used normal marginal distributions to generate the data but it was highly sensitive to the change of marginal distribution to gamma distribution\footnote{The numerical estimations from LNC are computed with $k=5$ (k being the number of nearest neighbors) and the default value of $\alpha$ parameter, using the toolbox available online at \cite{LNC}.}. The absolute mutual information error obtained using the LNC method was nearly an order of magnitude larger for the gamma function marginal distribution compared to the Gaussian marginals, for the same copula function. 
This result shows that the LNC method was strongly affected by the form of the marginal distribution, especially in the strongly correlated situations, e.g., large $r$ for Gaussian copula and small $\nu$ for student-t copula. In contrast to the LNC and the GC methods, the NPC had both desirable properties expected by an ideal estimator.

\begin{figure}[htp!]
  \includegraphics{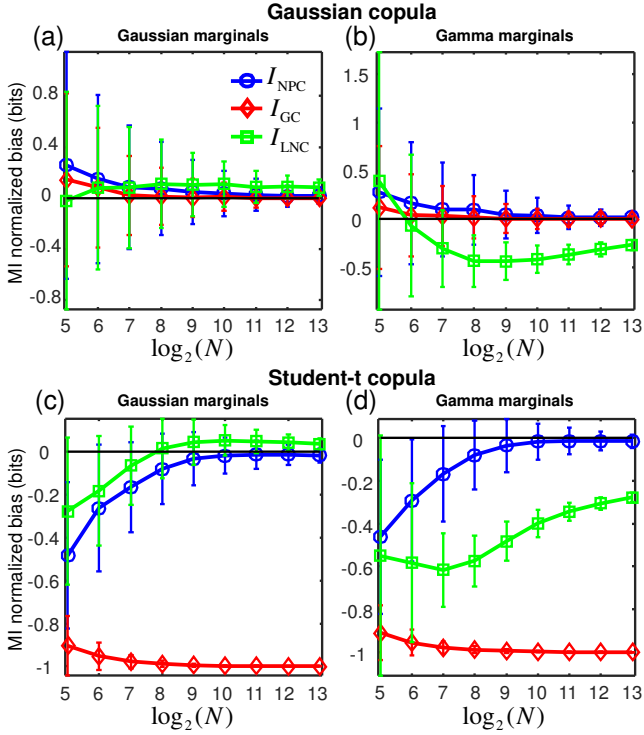}
  \caption{Mutual information bias is plotted for data similar to Fig.\ref{naiveI} but for different sample sizes ($N$) and the fixed values of $r=0.5$ and $\nu=0.5$. The errorbars represent standard deviation over data simulations.}
  \label{contNum_bias}
\end{figure}

First, it worked well for all the types of dependencies used to generate the data, giving low absolute errors for both data generated with the Gaussian and the student-t copula. Second, further quantification of the difference in the error in mutual information estimates when using either Gaussian or gamma marginal distributions (Fig.\ref{LNCdiff}) showed that the NPC-based estimator was not affected by the marginal distributions used to generate the data. The mutual information depends only on the copula, thus an ideal estimator should give equal results regardless of the marginal distribution. In sum, unlike previous methods the NPC-based estimator had the double advantage that it both functioned accurately for both types of copula families, including both linear and nonlinear dependency structures, and was insensitive to the marginal distributions.
\begin{figure}[t!]
  \includegraphics{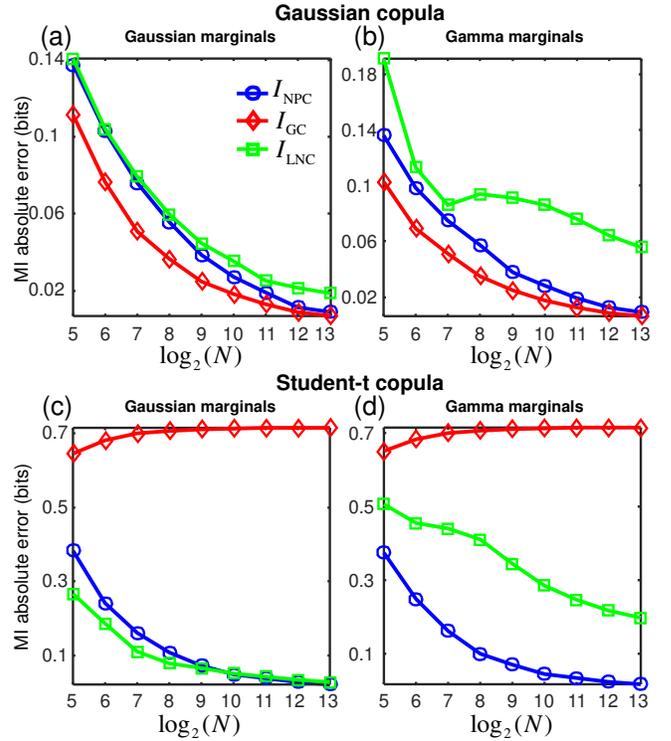}
  \caption{Mutual information absolute error similar to Fig.\ref{naiveI} but for different sample sizes ($N$), and the fixed values of $r=0.5$ and $\nu=0.5$.}
  \label{contNum}
\end{figure}
To further investigate the performance properties of the mutual information estimators, we computed the normalized bias and standard deviation of each of them, for the same data used in the above figures. The results (Fig.\ref{LNCcont_bias}) show that the better performance of the NPC estimator is largely due to a decrease in bias, but that the NPC estimator has also the additional desirable property of having in general less variance.
Given that, in practical applications, data available for information estimation are often scarce, it is important that an estimator is accurate also when small datasets are available. We thus investigated in Figs.\ref{contNum} and \ref{contNum_bias} how the performance of the NPC-based estimator varied with the sample size. We computed, for the four simulated data conditions and across a range of sample sizes ($N=2^5,\cdots,2^{13}$), the mutual information absolute error (Fig.\ref{contNum}) and the mutual information normalized bias (Fig.\ref{contNum_bias}). In these cases, we fixed the parameters of the copulas as $r=0.5$ for the Gaussian copula and $\nu=0.5$ for the student-t copula. The NPC method rapidly converged to a low error level with increasing sample size and had low error even at the smallest sample size. At most of the cases and sample sizes, the NPC method outperformed the LNC method, including for sample sizes as small as 64, for which there was an order of magnitude difference in the estimation error between the NPC and LNC methods for the simulated data with gamma function marginal distributions.
\subsection{Discrete variables}
We next considered the problem of estimating the mutual information between two random variables taking integer numerical variables. Having efficient information estimators in such cases is important for many applications. For example, in neuroscience experiments it is often important to estimate the information that the number of spikes emitted by neurons carry about sensory or behavioral variables  taking integer values. Note that, any discrete set of discrete variables can in principle be one-to-one mapped to a set of integer variables, with similar probability mass function of the original discrete variables; this makes the current setting quite general.

The local-likelihood kernel method requires a continuous, smooth, and integrable copula density, which is not the case for integer variables. We therefore used a simple approach to transform discrete data into the continuous domain, without affecting the information content, by adding appropriate noise to the data. This approach provided a single framework for computing mutual information between continuous and integer variables and their mixtures.
\subsubsection{Adapting the NPC estimator to discrete numerical variables}
We first examined how to transfer integer variables into the continuous domain without affecting the information content. Consider a bivariate set of integer variables $(n_X,n_Y)$. We can show that there exists proper noise variables $\epsilon_X$ and $\epsilon_Y$ independent from $(n_X,n_Y)$ such that  
\begin{equation}\label{noiseinf}
 I(n_X+\epsilon_X;n_Y+\epsilon_Y) = I(n_X;n_Y).
\end{equation}
One possible noise distribution satisfying Eq.(\ref{noiseinf}) is a union of uniform distributions filling the gaps between consecutive integer variables. Consider $\{n_i\}$ as the sorted set of integer variables ($n_i>n_{i+1}$ for all $i=1\cdots N_{\text{max}}-1$) according to their indices. We then add the following uniform noise 
\begin{equation}\label{noiseepss}
\epsilon_i\sim\mathcal{U}_{\left[ n_i,n_{i+1} \right] }
\end{equation}
to each integer $n_i$ transforming it to the corresponding $\tilde n_i$ in the real domain satisfying  $\tilde n_i>\tilde n_{i+1}$ for all $i=1\cdots N_{\text{max}}-1$. For $i=N_{\text{max}}$, we can define the noise as $\epsilon_{N_{\text{max}}}~\sim\mathcal{U}_{\left[ n_{i},n_{i}+1 \right] }$. We can then write the probability of the noised variable $\tilde n_i$ as
\begin{equation}
p(\tilde n_i)=p(n_i+\epsilon_i)= \sum_{n=1}^{N_{\max}} P(n)p(\epsilon_i=\tilde n_i-n)
\end{equation}

Since, based on the definition of the noise $\epsilon_i$, we have $p(\epsilon_i=\tilde n_i-n)=0$, for $n \neq n_i$ we will have
\begin{equation}\label{dec}
p(\tilde n_i)=  P(n_i)p(\epsilon_i).
\end{equation}
Similarly, the joint density can be decomposed as the product of the mass function of the integer variables $n_X$ and $n_Y$ and the noise densities
\begin{equation}\label{decc}
p(\tilde n_X,\tilde n_Y)=  P(n_X,n_Y)p(\epsilon_{n_X})p(\epsilon_{n_Y}).
\end{equation}
We then write the mutual information between the continuous variables $\tilde n_X$ and $\tilde n_Y$ as:

\begin{widetext}
\begin{eqnarray}
I(\tilde n_X;\tilde n_Y)&=&\iint\limits_{\tilde n_X\tilde n_Y} p(\tilde n_X,\tilde n_Y) \log_2 \frac{p(\tilde n_X,\tilde n_Y)}{p(\tilde n_X)p(\tilde n_Y)} d\tilde n_X d\tilde n_Y
\nonumber\\
&=&\sum_{n_X,n_Y}P(n_X, n_Y) \log_2 \frac{P( n_X, n_Y)}{P(n_X)P( n_Y)} \iint\limits_{\epsilon_{n_X}\epsilon_{n_Y}} p(\epsilon_{n_X})p(\epsilon_{n_Y}) d\epsilon_{n_X} d\epsilon_{n_Y}=I(n_X;n_Y)
\end{eqnarray}
\end{widetext}

\begin{figure}[htp!] 
  \includegraphics{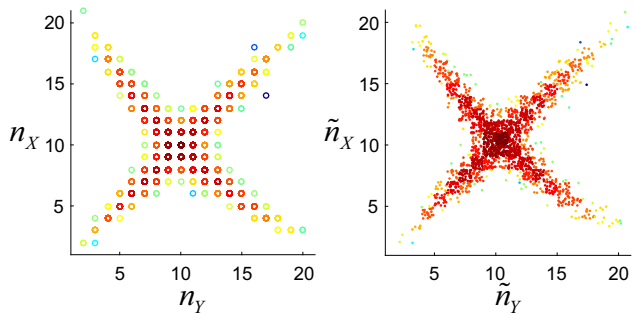}
  \caption{A set of bivariate integer data points (left) become continuous (right) after adding variables with proper noise distributions to each point.}
  \label{contin}
\end{figure}

which means that adding this noise and transforming the integer data to the real domain does not change the information between the variables. We can then use the variables $(\tilde n_X;\tilde n_Y)$ in the continuous domain together with the kernel copula to estimate their mutual information. An example simulation of such continuation of integer bivariate data into the continuous domain is shown in Fig.\ref{contin}. We note that a similar approach for continuation of the discrete domain into mixed variables for density estimation has been proposed in \cite{nagler2018generic}, to which we refer for further details.

\subsubsection{Testing the performance of the discrete NPC estimator}
To test the performance of the NPC method, we simulated data using Gaussian and student-t copulas with $r=0.5$ and $\nu=0.5$, respectively. Here, for the marginal distributions, we used Poisson distributions with a variable range of Poisson rates $\lambda=20,\cdots,70$ to see how changing the properties of the marginal distribution affects the mutual information estimation. Poisson distributions fit well many empirical data of relevance, such as the distribution of spike count of cortical neurons\cite{amarasingham2006spike}. We added noise to the data using Eq.(\ref{noiseepss}) and computed the corresponding copula and its entropy. We compared the NPC method with direct fitting using a Gaussian copula, because this comparison is useful to illustrate the specific advantages of a non-parametric copula. 
We also tested the NPC against the Pitman-Yor mixture (PYM)\footnote{For all the comparisons with PYM, we used the default setting of the codes available online at \cite{PYM}. We computed the joint entropy H(X,Y) from the multiplicities of all the unique pairs of integers in the data.} information estimation method~\cite{Archer2014}. We selected the PYM method for comparison because, as also confirmed by our experience on these simulated data, it has been shown~\cite{Archer2013,Archer2014} to further improve the performance of previous pioneering Bayesian estimators \cite{Nemenman2002,Nemenman2004}, and the latter compare favorably to other bias subtraction methods \cite{Nemenman2004,Panzeri2007}.
\begin{figure}[hbp!] 
  \includegraphics{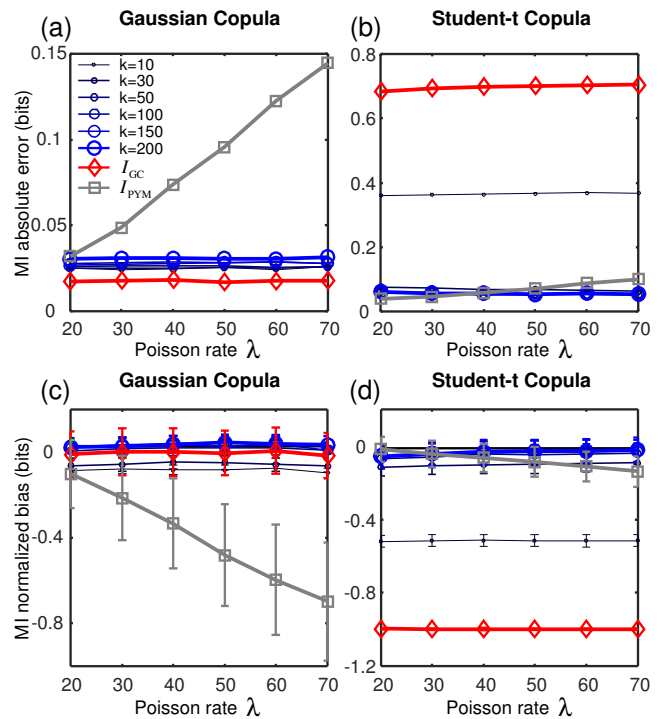}
\caption{The MI absolute error (top) and MI bias (bottom) are shown for the NPC model with a range of $k=10,\cdots,200$ number of bins, parametric Gaussian copula and PYM methods. To simulate the dataset, we used Poisson marginal distributions with $\lambda=50$ and the Gaussian (left) and student-t copulas (right) and generate $N=1024$ samples. The error bars of the MI bias plots are the standard deviation over 1000 data simulations.}
 \label{pymkall}
\end{figure}
 
\begin{figure}[b!]
  \includegraphics{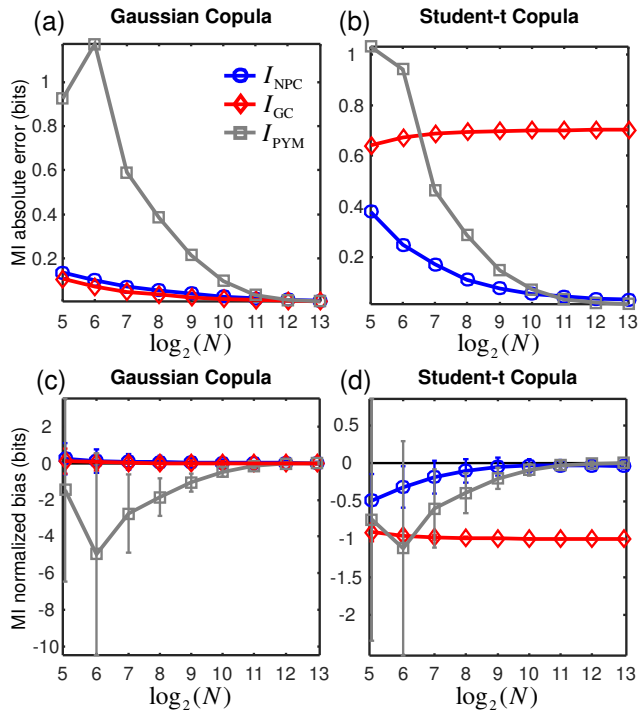}
  \caption{The MI absolute error (top) and normalized bias (bottom) are shown for different sample sizes for the NPC method, parametric Gaussian copula and PYM methods for Poisson marginal distributions with $\lambda=50$ and the Gaussian (left) and student-t copulas (right). The errorbars of the bottom panels are the normalized standard deviation computed over a set of data simulations.}
  \label{pymNum}
\end{figure}

We first focused on how to optimize the computation of the NPC estimator. As we did for the continuous case, we tested various values of $k$ (the binning parameter), compared models across simulation conditions, and analyzed estimation errors and biases as a function of sample size. 
In the discrete cases, we used the method used in \cite{Onken2008,Onken2016} to compute the ground truth mutual information.

The NPC-based estimator had a low and flat error across a wide range of $k$ values as is shown in Fig.\ref{pymkall}, with similar levels of error for $k > 10$. Also, the performance of the NPC estimator was insensitive to the properties of the marginal distributions and had similar levels of error across all tested values of $\lambda$. Further, the NPC estimator performed similarly well on both the Gaussian copula and student-t copula datasets. These results indicate that the NPC-based estimator performed similarly on integer variables as it did on continuous data. They also show flat normalized bias over the change of the Poisson rates (Fig.\ref{pymkall}). We then compared the NPC to other approaches over a range Poisson rates. As shown in Fig.\ref{pymkall}, the NPC estimator had significant advantages. Direct fitting with the GC approach worked well on the data generated from the Gaussian copula, but performed poorly on the data simulated with the student-t copula, as expected. The PYM approach performed worse than the NPC estimator on both cases, especially on the Gaussian copula case. The PYM method showed a strong dependency on the form of the marginals and had an order of magnitude larger errors for the largest values of $\lambda$. The NPC-based method was the only approach that generalized well across values of the marginal distributions and across the type of dependency structure in the data. Furthermore, the sample size dependency of different methods are shown in Fig.\ref{pymNum}. The performance of the NPC-based method, with a fixed Poisson rate at $\lambda=50$, had a weaker dependence on the sample size than the PYM method and had significantly lower estimation absolute error than the PYM method for sample sizes $N < 2^{10}$. Furthermore, the NPC shows small and flat normalized biases and variances over the same range of sample sizes, contrary to the PYM estimator which shows large negative normalized biases and large variances for small samples sizes.

\begin{centering}
\begin{figure}[t!] 
  \includegraphics{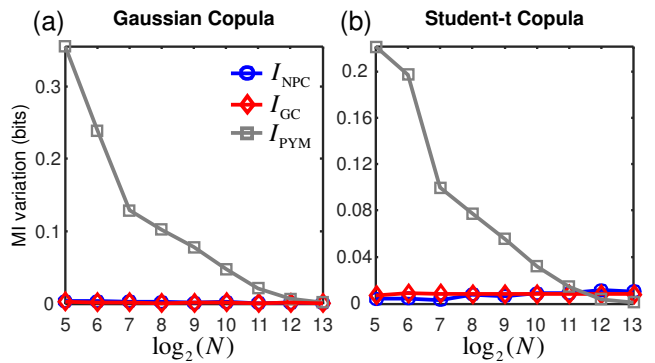}
  \caption{The standard deviation of MI is computed over a set of marginal distributions with firing rates $\lambda=20,\cdots,70$. The Poisson marginals are combined with (a) the Gaussian copula and (b) the student-t copula to generate the samples.}
  \label{pymvar}
\end{figure}
\end{centering}

These results further demonstrate that the NPC method has an important property of mutual information estimators, namely that they estimate similar mutual information values for a fixed dependency structure over a wide range of marginal distributions and sample sizes. In order to quantify the degree of the dependency of each estimator to the parameters of the marginal distributions, after fixing the dependency structure, we computed the variability in the estimated mutual information, measured as the standard deviation of the information values estimated over the a range of Poisson rates $\lambda$ (Fig.\ref{pymvar}). Across a wide range of sample sizes, the variability in the information estimate with varied $\lambda$ was flat for NPC and GC methods and low relative to that of the PYM method. The PYM shows strong marginal distribution dependency especially for smaller sample sizes. The NPC-based estimator therefore appeared unaffected by large changes in sample size or marginal distributions, consistent with what was observed in the continuous case. 

\section{Conclusions}
Here we developed a mutual information estimator based on non-parametric copulas. We have demonstrated that the method has several desirable features of a high-performance information estimator. First, the method is non-parametric, which means that assumptions about relationships in the data are not imposed. Second, the method is not sensitive to the distributions of individual variables (marginal distributions); rather, by virtue of its focus on the copula, it only takes into account the dependencies between variables. We were able to extend this advantage even to the discrete case, forming a single framework for the study of continuous, discrete, and mixed combinations of variables. Third, the NPC-based estimator worked well at low sample numbers, which has commonly been challenging for non-parametric approaches. We additionally demonstrated that this approach performed and generalized better than state-of-the-art mutual information estimators in many cases.

Many currently used mutual information estimators have made important progress in being able to estimate information accurately and from limited samples, also in cases when the underlying probability distributions do not necessarily fit traditional parametric families of probabilities.  However, these existing non-parametric methods do not explicitly single out the copula as the only part of the joint distribution that is taken into consideration for mutual information estimates~\cite{Archer2014,Gao2015,Kraskov2004}. We showed that estimators such as the kNN-based estimators and the PYM estimator were sensitive to the properties of the marginal distributions and can thus lead to inaccurate information estimates. For example, even with the same dependency structure and thus identical mutual information, these methods could erroneously estimate different levels of mutual information due to differences in the properties of the marginal distributions. By making use of copulas, we isolated the part of the joint distribution that is relevant for the mutual information and avoided contamination of the information estimates from irregularities in the marginal distributions. Both in the continuous and integer domains, the NPC estimator provided a stable information estimate across values of the marginal distributions and across sample sizes, and it shows less performance degradation at small sample numbers. These results indicate that the NPC approach is able to identify the dependency structure, which is exactly the property critical for the mutual information between the variables of interest, and the method was correctly not affected by changes in other aspects of the data. 

To model the copula, we made use of non-parametric methods. Contrary to parametric methods, non-parametric methods do not make strong assumptions with respect to the shape of the distribution and the dependency structure of the data. Here we showed that the use of non-parametric approaches allowed for successful information estimation both in data generated from Gaussian dependencies with linear correlations and from student-t copulas with only nonlinear relationships. In particular, we used the probit transformation in conjunction with principal component analysis to transform the data samples in the copula domain into a space that lends itself well to kernel density estimators. We made progress in kernel-based methods for copula density estimation. In such methods, the selection of the appropriate kernel bandwidth is a crucial factor for achieving faithful density estimates~\cite{Nagler2017}. We derived analytical solutions for the likelihood-estimated copula density with Gaussian kernels, making possible quick calculations of the density and the associated mean integrated square error. This allowed us to apply efficient methods for selecting the right kernel bandwidth. While other non-parametric copula methods such as splines smoothing~\cite{Kauermann2013} and Bernstein polynomials~\cite{Janssen2014} have been put forward, a recent comparison suggests that probit-transformation-based methods tend to outperform alternative non-parametric estimators over a wide range of used cases~\cite{Nagler2017} when combined with the local-likelihood density estimation \cite{Geenens2014}. 

Thus, the advantages of the NPC estimators result from being able to combine, for the first time into a single formalism, the best of two complementary approaches: the advantage of the copula to focus specifically on the parts of the probability distribution that are important for information and the advantage of non-parametric methods in being able to adapt to a wide range of situations. 

We tested the NPC-based estimator only in the bivariate case. The extension of copulas to multivariate cases has been developed through the vine-copula structures, showing that density estimators based on the vine-copula have better bias and variance scaling properties in terms of sampling size with respect to conventional non-copula based methods \cite{Aas2009,acar2012beyond,Geenens2014,Joe2015,NAGLER201669,Nagler2017}. Because the multivariate d-dimensional structures can be built using $d(d-1)/2$ bivariate copulas, the performance of the bivariate NPC suggests that similar trends are expected in higher dimensions. Investigation of the vine copula as a mutual information estimator in higher dimensions is an important area of focus for future work. 

We anticipate that, due to their adaptability to complex structures and their robustness to sample size, the NPC-based information estimator will be generally applicable in a wide range of fields and will advance and enhance the impact of information theory in many domains, in particular, application of information theory especially to biological problems in which data collection is constrained by insurmountable practical reasons and is both limited by the difficulty of estimating information accurately from limited samples~\cite{Brown2004,quiroga2009} and by the presence of complex nonlinearities ~\cite{franke2016structures}.

  As an important example, in neuroscience, hypotheses about how neurons encode information about certain behavioral variables (such as the parameters quantifying the nature of sensory stimuli or of behavioral choices) have thus far been limited to testing simple quantifications of the neural response, such as the number of action potentials fired in a given time window. Yet, evidence suggests that information may be encoded by more complex neural variables that include, for example, the pattern of firing of single neurons \cite{zuo2015complementary} or of neuronal populations \cite{runyan2017distinct}, or the interactions between the timing of action potentials and of continuous neural response variables such as the power or phase of brain oscillations~\cite{kayser2009spike}. The nature of the interactions between such neural variables and potentially complex external variables of ethological interest (such as the value of sensory stimuli of naturalistic complexity) is largely unknown and cannot be safely described by parametric methods. Yet, the number of samples that can be collected is limited by factors such as the small length of time in which a subject can perform a cognitive task. Our NPC information estimator can be used to measure accurately relationships between such neural and behavioral variables, helping researchers to crack the code used by neurons to mediate complex behaviors. The Matlab package implementing the pairwise local-likelihood copula and the NPC information estimation algorithm is available at \url{github.com/houman1359/NPC_Info.}

\begin{acknowledgements}
We thank members of our laboratories for helpful discussions, and Daniel Chicharro and Selmaan Chettih for feedback on the manuscript.
This work was supported by a Burroughs-Wellcome Fund Career Award at the Scientific Interface, the New York Stem Cell Foundation, and NIH grants from the NIMH BRAINS program (R01 MH107620), NINDS (R01 NS089521), and the BRAIN Initiative (R01 NS108410 and U19 NS107464) and the Fondation Bertarelli. C.D.H. and S.P. gave equal senior author contribution.
 \end{acknowledgements}

\bibliography{biinfo}

\end{document}